\documentclass{an}
\usepackage{graphicx}
\usepackage{times}
\Volume{00}              
\Year{0000}              
\Month{0}               
\Pagespan{1}{4}      

\begin{document}
\lhead[\thepage]{M.G. Watson {\it et al.}: The XMM-Newton Serendipitous Source Catalogue}
\rhead[Astron. Nachr./AN~{\bf XXX} (200X) X]{\thepage}
\headnote{Astron. Nachr./AN {\bf 32X} (200X) X, XXX--XXX}

\title{The XMM-Newton Serendipitous Source Catalogue}

\author{M.G.~Watson\inst{1}, J.P.~Pye\inst{1}, M.~Denby\inst{1}, J.P.~Osborne\inst{1},~
D.~Barret\inst{2},
Th.~Boller\inst{3},
H.~Brunner\inst{3,4},
M.T.~Ceballos\inst{5},
R.~DellaCeca\inst{6},
D.J.~Fyfe\inst{1},
G.~Lamer\inst{4},
T.~Maccacaro\inst{6},
L.~Michel\inst{7},
C.~Motch\inst{7},
W.N.~Pietsch\inst{3},
R.D.~Saxton\inst{1,8},
A.C.~Schr\"{o}der\inst{1},
I.M.~Stewart\inst{1},
J.A.~Tedds\inst{1},
N.~Webb\inst{2}
}

\institute{Department of Physics \& Astronomy, University of Leicester,
Leicester LE1~7RH, UK
\and
Centre d'Etude Spatiale des Rayonnements, F-31028 Toulouse, France
\and
Max-Planck-Institut f\"{u}r Extraterrestrische Physik,
85741 Garching, Germany
\and
Astrophysikalisches Institut Potsdam, D-14482 Potsdam, Germany
\and
Instituto de Fisica de Cantabria (CSIC-UC), 39005 Santander, Spain
\and 
Osservatorio Astronomico di Brera, 20121 Milano, Italy
\and
Observatoire Astronomique de Strasbourg, 67000 Strasbourg, France
\and
XMM-Newton SOC, VILSPA, 28080 Madrid, Spain
}
\date{Received {\it date will be inserted by the editor};
accepted {\it date will be inserted by the editor}} 

\abstract{We describe the production, properties and scientific
potential of the {\em XMM-Newton} catalogue of serendipitous X-ray sources.
The first version of this catalogue is nearing completion and is
planned to be released before the end of 2002.
}
\correspondence{mgw@star.le.ac.uk}

\maketitle

\section{Introduction}
Serendipitous X-ray sky surveys have been pursued with most X-ray
astronomy satellites since the {\em Einstein} Observatory. The
resultant serendipitous source catalogues have made a significant
contribution to our knowledge of the X-ray sky and our understanding
of the nature of the various Galactic and extragalactic source
populations. 
The {\em XMM-Newton} Observatory (Jansen {\it et al.}, 2001) provides
unrivalled capabilities for serendipitous X-ray surveys by virtue of
the large field of view of the X-ray telescopes with the EPIC X-ray
cameras (Turner {\it et al.}, 2001; Str{\" u}der {\it et al.}, 2001), and the
high throughput afforded by the heavily nested telescope modules. This
capability ensures that each {\em XMM-Newton} observation provides
significant numbers of previously unknown serendipitous X-ray sources
in addition to data on the original target (Watson {\it et al.},
2001). The compilation of a high quality serendipitous source
catalogue from the {\em XMM-Newton} EPIC observations is one of the
major responsibilities of the {\em XMM-Newton} Survey Science Centre
(SSC; Watson {\it et al.}, 2001).

This paper describes the approach taken to the production of the
catalogue and an overview of its main properties together with some
comments on its scientific potential. It should be stressed that we are
describing here the first installment of the catalogue; further releases are
planned as the constituent observations become public. The catalogue is
not yet complete; we are aiming for release by the
end of 2002. The description provided here thus relates to the
``working" version from which the catalogue will be extracted once
the full range of checks on integrity and quality have been completed.
Nevertheless the catalogue project is at a sufficiently mature stage
for the current version to be reasonably representative.

\section{Catalogue production}

\subsection{Selection of catalogue fields}
For the first installment of the catalogue we have selected {\em XMM-Newton}
observations (``fields") which fulfill the following criteria:
\begin{itemize}
\item observation made before {\em XMM-Newton} revolution 300 (to match
our planned catalogue release date to public availability of the
observations);
\item at least one EPIC camera having a
net exposure time  $\ge 1000$ seconds;
\item for the EPIC MOS cameras: data taken in any of the normal observing
modes; the outer 6 MOS CCDs in each camera normally accumulate images
even when the central CCD is taking data in window mode or timing
mode;
\item for the EPIC pn camera: data taken in full-frame and large window
imaging modes only; other EPIC pn modes either provide
very limited sky coverage (e.g. small window modes) or no imaging
data.
\end{itemize}
No other selection of fields, e.g. in terms of sky location, exposure
time etc., has been made. Fig.1 shows the sky distribution of
the catalogue fields. The overall sky distribution is
reasonably uniform, although there are some biases such as the paucity
of fields in the Cygnus region due to {\em XMM-Newton} visibility
constraints. 

\begin{figure}
\resizebox{\hsize}{!}
{\includegraphics[]{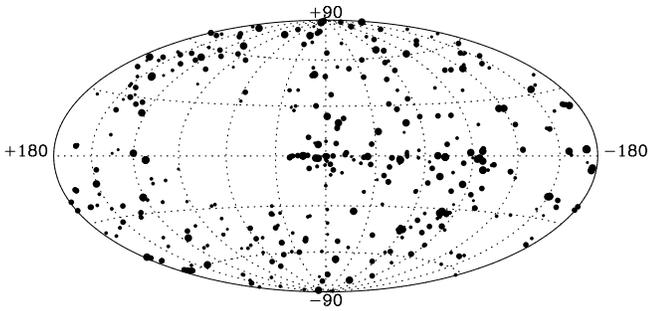}}

\caption{Sky distribution of the {\em XMM-Newton} fields in the catalogue shown in
Galactic coordinates. Size
of the symbols reflects the total number of serendipitous sources
detected: the average number is $\sim50$ per field.}

\label{f1}
\end{figure}

The current working version of the catalogue described in this paper
contains a total of around 700 fields. The average exposure time per
field is $\sim 20$ ksec for the EPIC MOS cameras and $\sim 15$ ksec
for the EPIC pn cameras. Only $\sim 67$\% of the total have data from
all 3 EPIC cameras. This surprisingly low fraction reflects the
exclusion of EPIC pn exposures not taken in full-frame or large
window mode, coupled with the fact that the planned observation with
all EPIC cameras did not take place due to operational problems for a
fraction of the observations. The fraction of total fields with data
from both EPIC MOS cameras is $\sim 85$\%.

\subsection{Data processing} 
\label{proc}
Data processing for the production of the {\em XMM-Newton} catalogue is
based closely on the standard SSC pipeline (see, e.g.,
{\tt http://xmmssc-www.star.le.ac.uk/}) used to process each
{\em XMM-Newton} dataset for distribution to the observer, and
population of the {\em XMM-Newton} Science Archive. Catalogue
processing uses a fixed software and calibration data configuration in
order to guarantee uniformity. The main processing stages for each
{\em XMM-Newton} observation are:
\begin{itemize}
\item production of calibrated events from the ODF science frames;
\item generation of the appropriate low-background time intervals
using a threshold optimized for point source detection;
\item generation of multi-energy-band X-ray images and exposure maps from the
calibrated events;
\item a four-stage source detection and parameterization procedure:
\begin{itemize}
	\item generation of a preliminary source list using a
	sliding-box technique and local background estimation; 
        \item
	generation of a background map from 2-D spline fits to the
	images with bright sources excised; 
        \item generation of a
	refined source list again using a sliding-box technique but
	employing the background map; 
        \item maximum-likelihood fitting
	and parameterization of sources in the refined list.
\end{itemize}
\item merging of the 3 camera-level source lists into an EPIC-level
source list with merging on the basis of positional coincidence alone;
\item cross-correlation of the source list with a variety of archival
catalogues and other resources using the CDS facilities in Strasbourg.
\end{itemize}
The source search approach utilized involves simultaneous fitting of 5
energy band images for each EPIC camera, thus producing a source list
for each camera which contains source and detection information in
each energy band (as well as the total band). The camera lists
combined in the penultimate stage described above thus produce a
merged source list which forms the reference source list for that
observation.

For the purposes of producing the {\em XMM-Newton} catalogue a number
of enhancements to the standard processing have been introduced (these
features will shortly be introduced into the standard processing
pipeline). The most significant of these are:
\begin{itemize}
\item inclusion of the EPIC pn ``out-of-time" events in the background
model, thus reducing the spurious detection rate significantly;
\item correction of the astrometric reference frame of each {\em XMM-Newton}
field using cross-correlation of the {\em XMM-Newton} source list with
the USNO A2.0 catalogue (see section \ref{ast}).
\end{itemize}

The current catalogue focuses on point-like sources. A search for
extended sources is included in the catalogue processing on an
experimental basis, but the results will not be included in the first version.
Experience with this analysis will form the basis for adding
extended source information to subsequent releases.
\subsection{Quality control}
\label{qual}
Although the source detection algorithm described above is now mature,
typically producing reliable source lists from most {\em XMM-Newton}
observations, the approach is not perfect and is known to have
problems in producing reliable results in a number of (rare)
circumstances. These include:
\begin{itemize}
\item where the EPIC image contains a bright point-like
or high surface brightness extended source, small errors in the PSF
model or 
background estimation can lead to significant numbers of spurious
sources being detected at the bright source periphery;
\item the edges of the EPIC CCDs (primarily in MOS1) can show occasional
brightening leading to spurious detections.
\end{itemize}
Each {\em XMM-Newton} field included in the catalogue is therefore
visually screened to locate such defects. Where problems are noted the
sources affected are ``flagged" and the flag values transferred to the
source lists. In rare cases ($<10$\% of the total) the entire field
has significant problems, e.g. very high background or very high
surface brightness diffuse sources, which mean that it is of marginal
value for detecting serendipitous sources. Such fields will be
excluded from the final catalogue. Apart from these rare cases, the
median fraction of sources flagged as being spurious amounts to only
4\% overall, reflecting the maturity of the source detection approach.
The screening process does not address the reality of low significance
detections (i.e. sources are {\bf not} flagged simply because they
appear to be marginally significant visually); this issue is being
pursued via simulations.

\section{Catalogue properties}
\label{catprop}
\subsection{Source numbers}
The working catalogue contains a total of $\sim 78000$ 
source detections in any EPIC camera (i.e. in one or more cameras) and
a total of $\sim 11000$ sources detected in all 3 EPIC cameras. These
numbers refer to a broad-band detection above a likelihood of 10,
corresponding to $\approx 4\sigma$. At this significance the {\it a
priori} probability of spurious detections is low,
corresponding to $<1$ spurious source per field, although simulations
are underway to verify the calibration of the likelihood
parameterisation. The total sky area
covered for detections in any camera is $\sim 130$ sq.deg., whilst for
detection in all 3 cameras the area is $\sim 90$ sq.deg.

The large difference between the total number of sources detected and
the number detected in all three cameras is due a combination of three
effects: the fact that only 2/3 of the fields have data from all three
cameras, because the fields-of-view of the cameras do not
precisely overlap and because source lists at a likelihood of 10
are significantly incomplete.

For the released version of the catalogue these numbers will be
reduced by the fraction flagged as spurious (and the small number of
fields totally excluded). We also anticipate setting a somewhat higher
likelihood threshold once we have completed our investigation of the
reliability of detections as a function of likelihood, being pursued
by simulations.

\begin{figure}[h!]
\begin{center}
{\includegraphics[height=7cm]{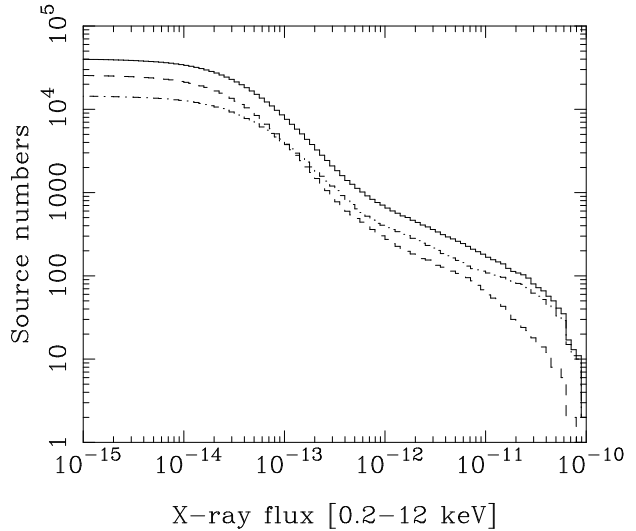}}
\caption{
Uncorrected $\log N-\log S$ distribution for all EPIC pn sources
in the working catalogue. The solid curve is for all sources, the
dashed curve is for high latitude sources ($|b|>20^\circ$)  
and the dashed-dot curve for low
latitude sources ($|b|<20^\circ$).}
\label{f2}
\end{center}

\end{figure}

\subsection{Source count distribution}
\label{scnt}
Figure 2 shows the $\log N -\log S$ distribution for all EPIC
pn sources in the working catalogue. The distribution is {\bf not}
corrected for sky coverage, i.e. how the actual sky area covered varies with 
X-ray flux. Comparing the uncorrected $\log
N -\log S$ with the expected source counts demonstrates that the
catalogue is essentially complete down to an X-ray flux $f_X \approx 4\times
10^{-14}\mathrm{\ erg\ cm^{-2}\ s^{-1}}$ (0.2-12 keV) (equivalent to
$f_X \approx 2\times
10^{-14}\mathrm{\ erg\ cm^{-2}\ s^{-1}}$ in the 0.5-2 keV band and 
$f_X \approx 8\times
10^{-15}\mathrm{\ erg\ cm^{-2}\ s^{-1}}$ in the 2-10 keV band). This
limit is
in line with
expectations given the exposure time distribution of catalogue fields.
Around 30\% of the catalogue sky area is covered to $f_X \approx
10^{-14}\mathrm{\ erg\ cm^{-2}\ s^{-1}}$ (0.2-12 keV). Work to establish the
definitive coverage corrections is underway and will be part of the
ancillary information included with the released version of the
catalogue.

\subsection{Count rate comparison}
\label{crate}
For the sources detected in more than one EPIC camera we can compare
the count rates to investigate the accuracy and reliability of the
catalogue (as well as the underlying calibration data). 
The overall count rate
fidelity found is excellent. Investigation of the relatively small
fraction of outliers has shown that these mostly correspond to
spurious detections which fortuitously line up with detections in
another camera. These and other anomalies will be removed in the
released version of the catalogue.

\subsection{Astrometry}
\label{ast}
For each {\em XMM-Newton} field, the catalogue processing (section 
\ref{proc}) attempts to correct the astrometric reference frame using
cross-correlation of the {\em XMM-Newton} source list with the USNO
A2.0 catalogue (Monet {\em et al.}, 1998). The technique employed involves finding the maximum
likelihood in a grid of trial astrometric shifts and rotations with a
likelihood function depending on the angular separation between each
potential {\em XMM-Newton}--USNO object match. If an acceptable
solution is found ($>70$\% of fields) the resultant astrometric
correction is applied to the {\em XMM-Newton} source list for that
field. (The cases where an acceptable solution is not found are
primarily fields with low numbers of X-ray sources and/or fields with
high optical object density).

The results of applying this technique to the catalogue fields can be
employed to quantify the initial accuracy of the astrometry of each
{\em XMM-Newton} field (i.e. before correction). 
Fits to the distribution of shifts in RA, Dec and field rotation imply
that the intrinsic accuracy of the {\em XMM-Newton} field astrometry
(as determined solely from the in-orbit attitude solution) can be
characterized by a Gaussian with $\sigma \approx 1.3$
arcsec. {\it After} correction using this technique, the residual
field astrometric errors are of the order 0.5-1 arcsec, close to the
nominal 1 arcsec. astrometric accuracy of the USNO catalogue itself.

As the typical {\it statistical} error-circle for a faint {\em
XMM-Newton} source has $\sigma_\mathrm{stat}\approx 1-2$ arcsec., the
size of the field systematic component determined justifies the
$\sim5$ arcsec. positional accuracy which has been assumed to date as
the effective $\sim 90$\% confidence radius of uncorrected positions,
e.g. for the identification of {\em XMM-Newton} source counterparts.

\section{Scientific potential of the catalogue}
The {\em XMM-Newton} catalogue represents a significant resource that
can be used for a variety of astrophysical projects. Although deep
Chandra and {\em XMM-Newton} pencil-beam surveys (e.g. 
Mushotzky {\it et al.}, 2000; Hasinger {\it et al.} 2001; Giacconi
{\it et al.} 2001; Brandt {\it et al.} 2001) have probed the faintest
parts of the extragalactic source population, the {\em XMM-Newton}
catalogue also can make a major contribution. The {\em XMM-Newton}
catalogue reaches modest depths ($f_X
\approx 10^{-14}\mathrm{\ erg\ cm^{-2}\ s^{-1}}$) at high coverage
(tens of square degrees). As this flux limit is where
the bulk of the objects that contribute to the X-ray background lie
(due to the fact the $\log N -\log S$ distribution breaks to a flatter
slope at around this flux), the large samples of medium-deep flux
sources that the {\em XMM-Newton} catalogue provides will thus be a
significant resource
for X-ray background studies.

\begin{figure}
\begin{center}
{\includegraphics[height=7.5cm]{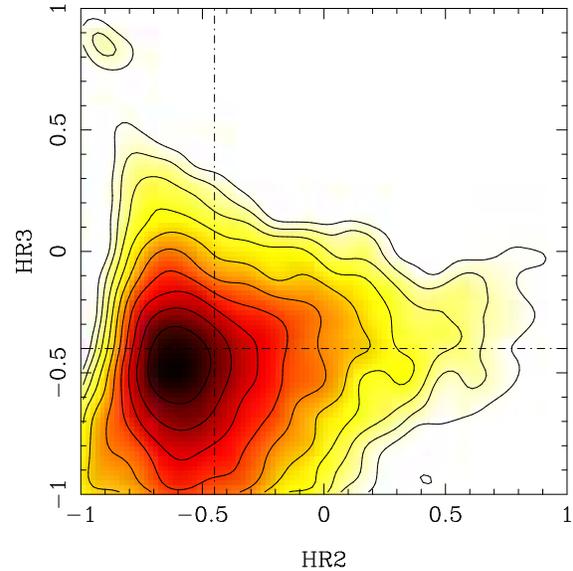}}
\caption{X-ray colour-colour plot showing the distribution of HR2-HR3
values for EPIC pn catalogue sources ($|b|>20^{\circ}$ only). The contours
indicate the logarithmic density of points
down to $\sim 1$\% of the peak. The dashed-dot lines indicate the
approximate HR2, HR3 values for the X-ray background. HR2 is the ratio
(B-A)/(A+B) and HR3 is the ratio (C-B)/(B+C) where A= 0.5-2 keV count rate, 
B=2-4.5 keV count rate and C=4.5-7.5 keV count rate.} 
\end{center}

\label{f5}
\end{figure}

The {\em XMM-Newton} catalogue also provides a rich resource for
generating well-defined samples for specific studies, utilizing the
fact that X-ray selection is a highly efficient (arguably the most
efficient) way of selecting certain types of object, notably AGN,
clusters of galaxies, interacting compact binaries and very
active stellar coronae. AGN samples have obvious value in
evolution studies and cluster samples can provide, potentially, key
measurements of cosmological parameters. Selecting ``clean" samples
requires knowledge of the likely parameter ranges of different types
of object: some of this is already known but further ``calibration" of
this concept is one of the key aims of the SSC's XID programme (Watson
{\it et al.}, 2001; Barcons {\it et al.}, 2002).  The inclusion of
matches with archival catalogues for each {\em XMM-Newton} catalogue
source also provides a valuable starting point for investigation of
well-defined samples.

To illustrate some of the potential described above, Figure 4
shows the overall distribution of X-ray colours in the catalogue. This
figure provides a glimpse of the power of the {\em XMM-Newton}
catalogue for providing interesting samples. The spectrum of the X-ray
background corresponds to HR2 $\approx -0.45$, HR3 $\approx -0.4$, but
evidently the bulk of the catalogue sources have spectra softer than
this. Thus merely by extracting a subset of {\em XMM-Newton} sources
with X-ray colours harder than these limits, one automatically selects
those objects that must be an important constituent of the background.
Optical and near-IR observing programmes to investigate the nature of
the hard source samples constructed in this manner are already
underway.

As well as the potential for building up large samples, the brighter
sources in the {\em XMM-Newton} catalogue also provide the prospect of
obtaining high quality X-ray spectra and time series data (and
morphology). Amongst the catalogue sources more than 15\% have more
than 200 EPIC pn counts, enough for a reasonable spectral
characterization and crude variability indications, whilst 3\% have
more than 1000 counts, sufficient for a very good X-ray spectral
measurement or variability analysis.

\acknowledgements
Our ability to construct a high quality {\em XMM-Newton} catalogue
rests on the efforts of many members of the SSC, ESA science
operations (SOC) and {\em XMM-Newton} instrument teams who have
contributed to every aspect of the project from operating the
satellite, calibrating the instruments, writing the science analysis
software and constructing and operating the processing pipelines.

\end{document}